# Roman CCS White Paper

# Characterizing the Galactic population of isolated black holes

**Roman Core Community Survey:** Galactic Bulge Time Domain Survey

**Scientific Categories:** stellar physics and stellar types; stellar populations and the interstellar medium

**Additional scientific keywords:** astrometry, black holes, gravitational microlensing

**Submitting Author:** Casey Y. Lam (UC Berkeley)

**List of contributing authors:**
Natasha Abrams (UC Berkeley)
Jeff Andrews (University of Florida)
Etienne Bachelet (IPAC)
Arash Bahramian (Curtin Institute of Radio Astronomy)
David Bennett (NASA Goddard, UMd College Park)
Valerio Bozza (Salerno University)
Floor Broekgaarden (CfA | Harvard & Smithsonian)
Sukanya Chakrabarti (University of Alabama, Huntsville)
William Dawson (Lawrence Livermore National Laboratory)
Kareem El-Badry (Harvard)
Maya Fishbach (CITA, University of Toronto)
Giacomo Fragione (CIERA, Northwestern)
Scott Gaudi (The Ohio State University)
Abhimat Gautam (UCLA)
Ryosuke Hirai (Monash University)
Daniel Holz (University of Chicago)
Matthew Hosek Jr. (UCLA)
Macy Huston (Penn State)
Tharindu Jayasinghe (UC Berkeley)
Samson Johnson (NASA JPL)
Daisuke Kawata (University College London)
Naoki Koshimoto (Osaka University)
Jessica R. Lu (UC Berkeley)
Ilya Mandel (Monash University)
Shota Miyazaki (JAXA/ISAS)
Przemek Mróz (University of Warsaw)
Smadar Naoz (UCLA)

Clément Ranc (IAP, Sorbonne University)
Dominick Rowan (The Ohio State University)
Rainer Schödel (IAA-CSIC, Granada, Spain)
Tomer Shenar (University of Amsterdam)
Josh Simon (Carnegie Observatories)
Rachel Street (Las Cumbres Observatory)
Takahiro Sumi (Osaka University)
Daisuke Suzuki (Osaka University)
Sean Terry (UC Berkeley)

**Abstract:** Although there are estimated to be 100 million isolated black holes (BHs) in the Milky Way, only one has been found so far, resulting in significant uncertainty about their properties. The Galactic Bulge Time Domain Survey provides the only opportunity in the coming decades to grow this catalog by order(s) of magnitude. This can be achieved if 1) Roman's astrometric potential is fully realized in the observation strategy and software pipelines, 2) Roman's observational gaps of the Bulge are minimized, and 3) observations with ground-based facilities are taken of the Bulge to fill in gaps during non-Bulge seasons. A large sample of isolated BHs will enable a broad range of astrophysical questions to be answered, such as massive stellar evolution, origin of gravitational wave sources, supernova physics, and the growth of supermassive BHs, maximizing Roman's scientific return.

# Table of contents

## Summary of needs and recommendations to find black holes

In order to find black holes, the needs of the Galactic Bulge Time Domain Survey (GBTDS) beyond the notional design are:

- Precise relative astrometry (optimal 0.1 mas, minimal 0.3 mas) over the full field of view
- Minimization of Roman's observing gaps: placing as many seasons consecutively as possible; additional Bulge seasons beyond the notional 6 for the GBTDS (optimally have at least daily cadence)
- Continuous observations of the Bulge from the ground; data must be public with no proprietary period to enable follow-up; majority of observations optimally taken in the IR.

We thus make the following recommendations (a summary of the minimal and optimal strategies is summarized in Table 1):

- Roman observations during Bulge seasons that are current gaps in the notional GBTDS
- Ensuring relative astrometry is included as a driving science requirement during survey design and software/pipeline development
- Coordinating observations with ground-based IR facilities, in particular PRIME

## Black holes allow us to understand how the Universe works

Stellar-mass black holes (BHs) are an end product of stellar evolution formed when massive stars exhaust their fuel and gravitationally collapse. The number, mass function, velocity distribution, and binary fraction of BHs are key to understanding how they form, evolve, and interact. BH properties are an observational boundary condition of stellar evolution (Figure 1) needed to test the highly uncertain physics of massive star death, such as implosion/explosion mechanisms [1-4], natal kicks [5-9], and the types and importances of different binary interactions [10-13]. In turn, these are needed to understand chemical enrichment, gravitational wave sources, galaxy formation and evolution, and supermassive BH growth.

Despite their importance to a broad range of astrophysics, the basic properties of the Galactic BH population are almost entirely unknown. Although there are expected to be $10^7$ - $10^9$ BHs in the Milky Way with >80% of those being isolated [14-19], **all but one of the two dozen Galactic BHs with mass measurements reside in binaries** [20-27]. **This highly biased sample hinders our ability to understand massive stellar evolution and related astrophysics. Isolated BHs are also needed to understand the rapidly growing catalog of merging BHs detected via gravitational waves** [28]. Merging BHs are a rare outcome of stellar evolution, where the formation channel(s) are highly sensitive to uncertain physics and initial conditions [29-31]. Only a catalog of "ordinary" isolated BHs can avoid these evolutionary biases, probe the dominant BH formation channel(s), and contextualize the "extraordinary" population of merging BHs. **Gravitational microlensing is the only practical way to find and characterize isolated BHs** (Figure 2), with the first such measurement made last year [25-27].

Isolated BHs should be a driving science case in designing the GBTDS. First, they are key to understanding a broad range of astrophysics. At present, BHs are not part of Roman's main science objectives (cosmology and exoplanets); inclusion of BHs and compact objects maximizes the science return of the mission. Second, the majority of the observations needed are identical to those required for exoplanetary microlensing; a search for BHs with microlensing naturally fits into the same survey. Third, the observations will be of broad utility: in addition to finding BHs, they will improve our understanding of Galactic structure [32-35], characterize longer-duration IR transients [36], and enable proper motion/astrometry studies [37-40].

## Roman is the best way to find many isolated black holes

Roman's wide field of view and ability to simultaneously obtain precise photometry and astrometry can enable the GBTDS to grow the catalog of isolated BHs by orders of magnitude, given the current sample size of one. **No other facility in the next several decades can achieve this goal.** Ground-based surveys cannot achieve the requisite spatial resolution or depth; as they are seeing-limited, it is difficult to detect photometric microlensing parallaxes and impossible to measure astrometric microlensing signals, both of which are necessary to measure BH masses and velocities. Targeted follow-up to obtain astrometry with facilities like Keck + laser guide star adaptive optics or Hubble Space Telescope (HST) is observationally expensive, and thus is limited in the number of candidates that can be followed up [25-27, 41]; it can only grow the sample by a few over 5 years. HST cannot perform a GBTD-like survey because its field of view is 100x smaller, nor is it optimized for quick slew/settle, making it prohibitive to observe at a cadence needed to characterize a significant number of microlensing events [42]. And although Gaia has exquisite astrometry, it does not perform well toward the Bulge due to crowding and extinction [26, 43]; it can find at most a few isolated BHs [44].

The GBTDS can detect hundreds of BHs via microlensing. Given the notional GBTDS design, [45] predicted $3 \times 10^5$ microlensing events will be detected. BH lenses account for ~1% of all microlensing events [46, 47]; the other 99% comprise stars (which can host planets), white dwarfs, and neutron stars. Accounting for uncertainties in both Galactic models and the order-of-magnitude number of BHs, a very conservative lower estimate still predicts at least 300 BHs detected by Roman. Properly designed, the GBTDS will be able to measure the masses and velocities for a significant number of these BHs, which will grow the existing sample by orders of magnitude and help answer fundamental questions in stellar astrophysics, such as the magnitude of BH natal kicks [48] and the initial-final mass relation of massive stars [49].

## Enabling Roman to uncover the isolated black hole population

### Differences between finding exoplanets and black holes with microlensing

The design of the GBTDS needed to meet the exoplanet detection requirements also make it excellent for finding BHs with microlensing. The "non-negotiables" to meet the exoplanet requirements are a monitored area of ~2 $deg^2$, at least 6 x 60 day seasons, with a cadence of at least 15 minutes, and the longest possible total time baseline (see Appendix for a summary).

However, the strategies for finding BHs and exoplanets are not identical. First, because the lensing timescale is proportional to the square root of the lens mass, a microlensing event due to a $10 M_\odot$ BH lens is on average 100x longer than that due to a $1 M_J$ planetary lens (Figure 3).

BH events will not be completely observed by Roman due to the gaps in Bulge visibility (Figure 4). In addition, the fiducial GBTDS design has a ~2.5 year gap between the first three and last three Bulge seasons. Minimizing the gaps by having some observations during all Bulge seasons and placing as many seasons consecutively as possible can help mitigate this. Having ground-based observations when Roman cannot see the Bulge will also be crucial to characterizing BHs and modeling events with sparse lightcurve coverage. The PRIME telescope would be ideal, as it has the same detectors as Roman. However, synergies with other facilities from ground and space, including Rubin, JASMINE, and ULTIMATE-Subaru would be beneficial, as they are highly complementary to PRIME in depth and wavelength. The data from these synergistic surveys also need to be publicly available with no proprietary period, like the Roman data, to announce and enable follow-up by the community.

Second, the astrometric signals of BHs and exoplanets are very different. Like the timescale, the astrometric signal is proportional to the square root of the lens mass, so on average a $10M_\odot$ BH lens will produce an astrometric shift 100x larger than a $1M_J$ planetary lens. Although planetary astrometric shifts are too small to be detectable, those of BHs are detectable with Roman (Figure 5). Precise relative astrometry is crucial for BH microlensing studies as it enables their masses to be measured: it is impossible to measure lens masses with photometry alone because of a fundamental degeneracy between lens mass and proper motion. We note that astrometry is also important for the exoplanetary science cases, but the needs (e.g. measuring lens and sources separating several years after the event) are different than BHs.

## Survey design: trade space, metrics and figures of merit

Table 1 summarizes the minimal and optimal observational strategies for enabling BH searches with microlensing, given the bounds imposed on the GBTDS by the exoplanet requirement. It is very similar to the notional design; the additions are relative astrometry requirements. Figure 6 shows the number of characterizable BHs as a function of astrometric precision. The minimal strategy would yield a few 10s of BHs, while the optimal strategy would yield a few 100s.

Having as many Bulge seasons arranged consecutively as possible is critical for BH searches, and the ~2.5 year gap in the notional design needs to be filled. At minimum, observations with an astrometric precision of 0.3 mas (i.e. 10 exposures stacked, for single exposure precision of 1 mas) at a weekly cadence within each Bulge season are needed to properly sample the astrometry. In addition, concurrent observations of GBTDS fields with ground-based facilities will be crucial for photometric coverage of the full lightcurve; their Bulge observational duty cycle of ~75%, vs. Roman's ~40% is important for covering long-duration BH events.

Although increasing the survey area will linearly increase the number of BHs detected, this comes at the strong expense of characterization. If there is a choice between filling Bulge season observational gaps in existing fields vs. adding new fields at a sparse (>1 day) cadence, the former will be significantly better for the yield of characterizable BHs.

Finally, Table 2 lists synergies with other white papers and pitches that detail topics we only briefly mention in this white paper.

## Figure 1: A representative sample of black holes is needed to understand their formation and population properties

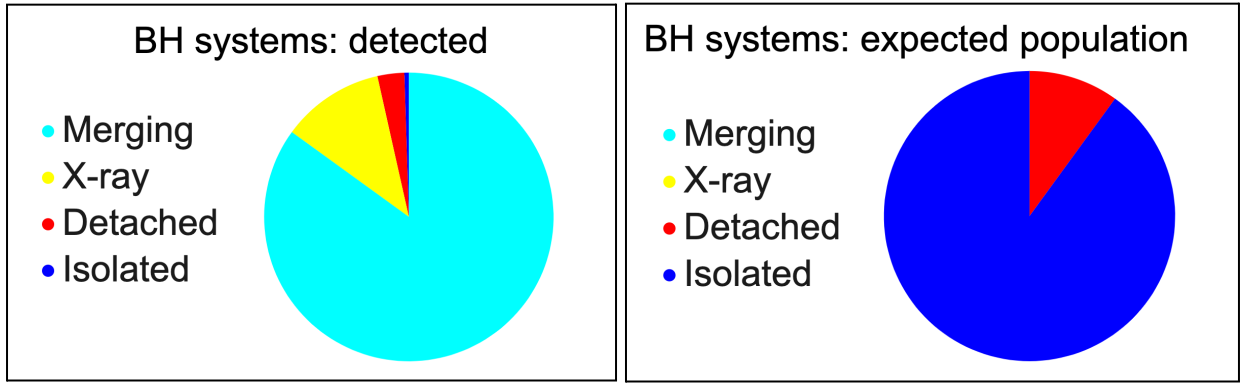

In order to understand massive stellar evolution, the stellar progenitor properties, environment, and evolution must be connected to the resultant BH. Presently, the impact of initial conditions and interactions on the resultant BH system are very uncertain. Although ~200 stellar-mass BHs have been detected to date, the vast majority are from extragalactic mergers detected via gravitational waves [28], which represent a very rare and exotic evolutionary channel. **Even the majority of the ~30 detected BHs in our own Milky Way and Magellanic clouds are in close and highly interacting X-ray binaries**, another rare outcome [20-24, 50-52]; **their properties cannot be used to infer those of the full Galactic BH population** [53-55].

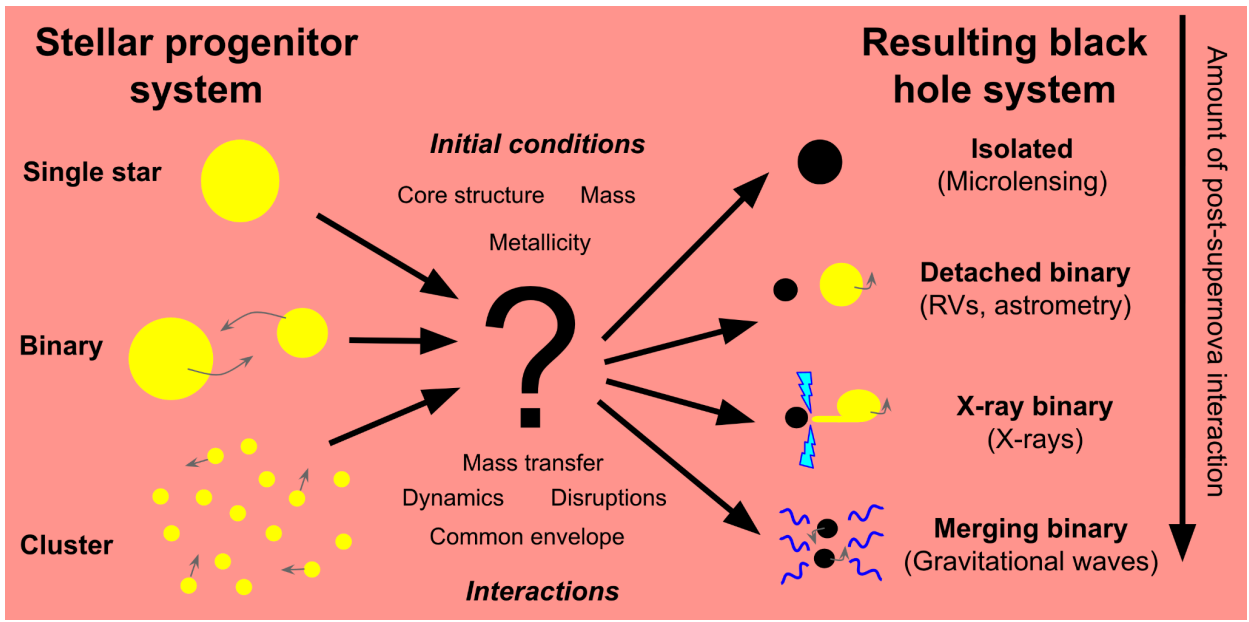

In addition, isolated BHs are more likely formed with less (or no) mass transfer phases, which can help with mapping their pre-supernova stellar properties and understanding supernova physics, and are also a unique probe of BHs in disrupted binaries. Roman can add up to a few hundred isolated BH mass measurements to this sample and ameliorate this issue.

## Figure 2: Gravitational microlensing can characterize isolated black holes

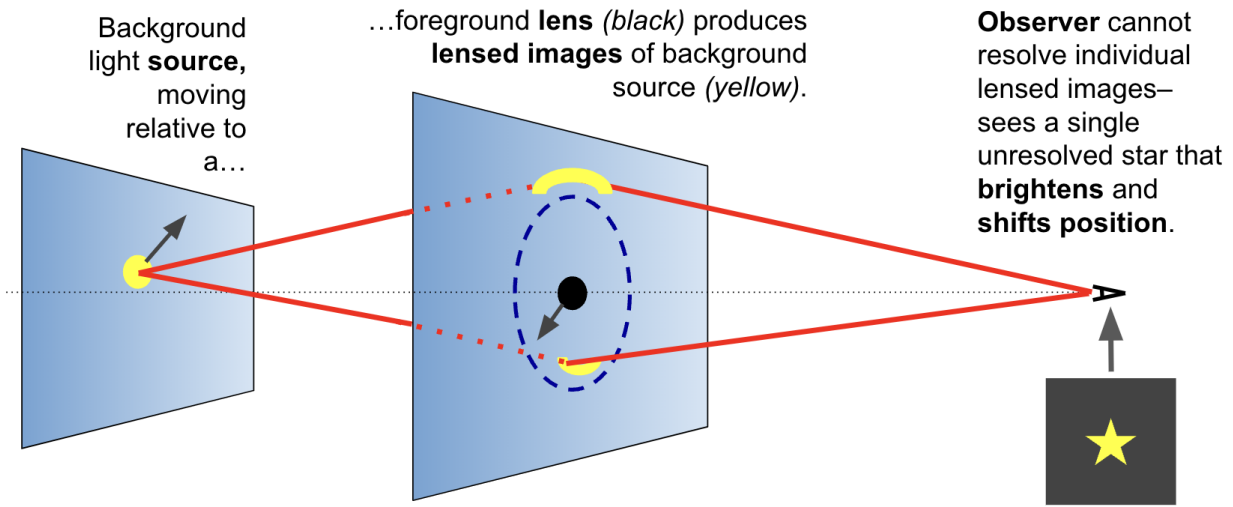

Background light **source,** moving relative to a…

…foreground **lens** *(black)* produces **lensed images** of background source *(yellow)*.

**Observer** cannot resolve individual lensed images– sees a single unresolved star that **brightens** and **shifts position**.

Gravitational lensing is the ideal way to detect isolated BHs, as the phenomenon depends on the mass, and not the luminosity, of the foreground lens. In the Milky Way, gravitational lensing occurs when a background source star (usually in the Bulge) and a foreground lens (star or compact object) coincidentally align along an observer's line of sight. The lens deflects the light from the background source and produces two lensed images.

For lens masses and distances typical of the Galaxy, the lensed images are separated by <1 mas, and hence are unresolvable in the optical/IR. As the background source and foreground lens move, align and separate, the observer instead sees a single unresolved star that appears to temporarily brighten (Figure 4) and shift position (Figure 5).

The mass and transverse velocity of the lens can be measured by jointly modeling both the brightening (photometric microlensing) and shift (astrometric microlensing). If the lens is >5M$_\odot$ and a star can be ruled out (e.g. the lens is very luminous), then the lens can be confirmed as a BH. For more details on the methodology, see [26]; for discovery papers, see [25, 27].

**Having astrometry is crucial in measuring the lens mass and velocity; with photometry alone, this is not possible** due to a fundamental degeneracy between the mass and velocity of the lens. Although there have been several excellent BH candidates over the past 20 years, they cannot be confirmed because there were no astrometric measurements [56-57].

For more background information on photometric and astrometric microlensing, see [58-62].

**Figure 3: Microlensing timescales of exoplanets vs. black hole lenses**

Schematic of the distribution of Einstein crossing times $t_E$ for exoplanets (blue) vs. BHs (black). For comparison, stellar lenses are shown in the dashed gray line. Note the logarithmic scaling on the x and y axes. **The typical lensing timescales of BHs are 100x longer than that of exoplanets.** Figure based upon a combination of [47, 63].

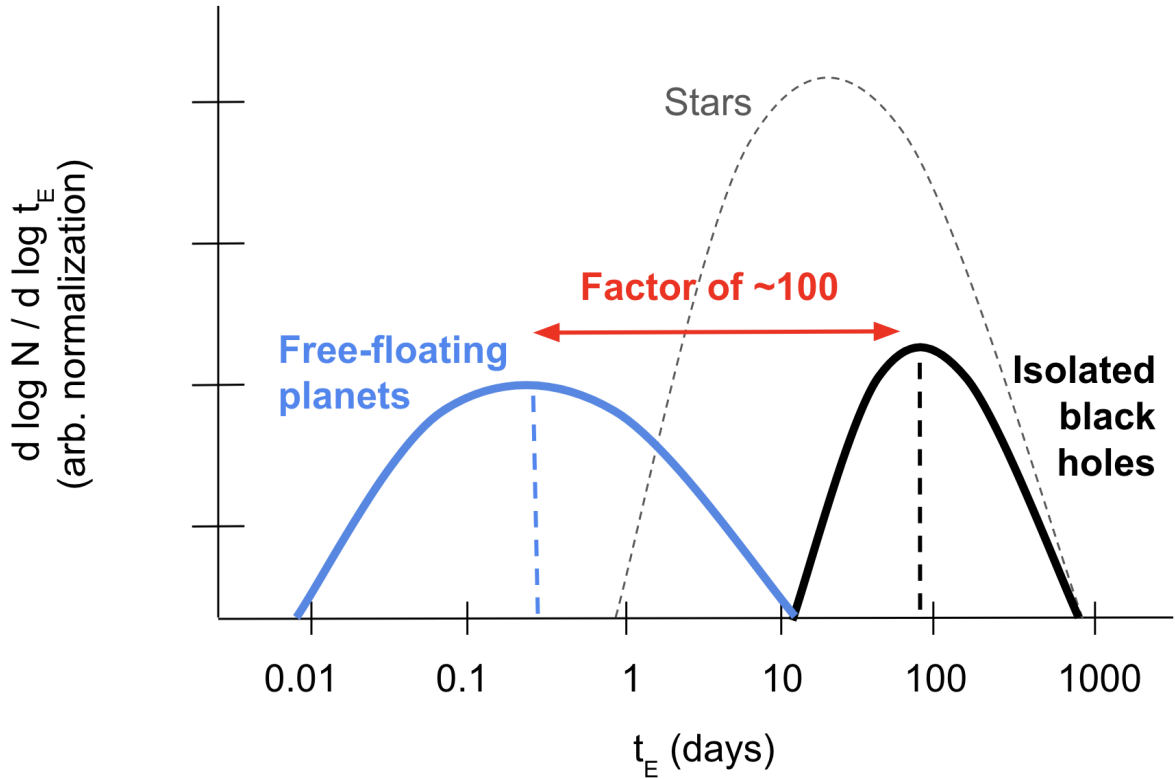

# Figure 4: Microlensing lightcurves of exoplanet vs. black hole lenses

Two examples of planetary microlensing lightcurves. Left: 0.5 $M_J$ planet separated 1.5 AU from its 0.25$M_\odot$ host star. Right: 1.8 $M_J$ free floating planet. Figures adapted from [45, 63]. The entire lightcurve can fall within a 72 day Bulge window.

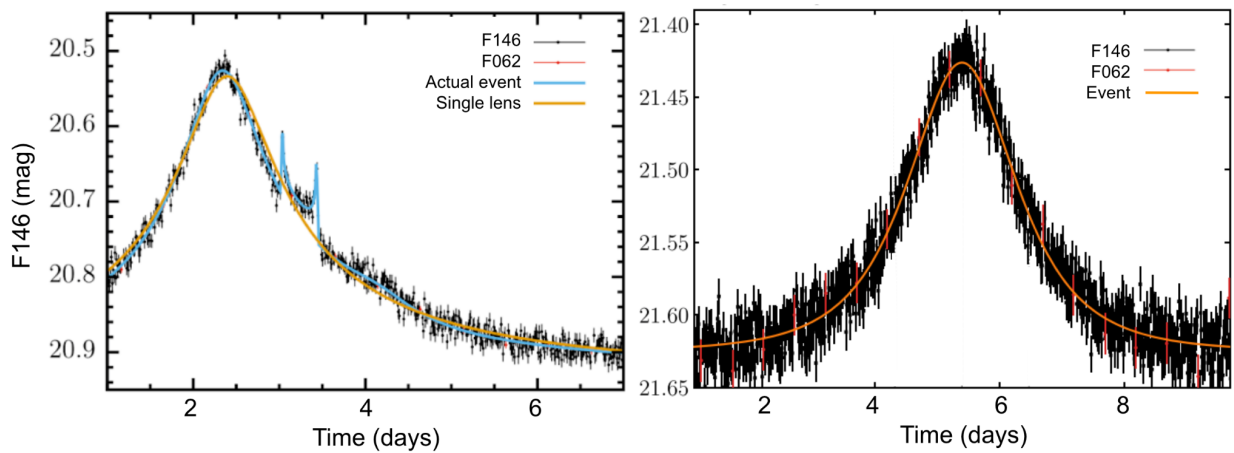

Example of a BH microlensing lightcurve. The lightcurve is sampled at the GBTDS notional/minimal strategy (left), GBTDS optimal strategy (middle), and a ground-based cadence (right). The uncertainties are representative of Roman [45] and PRIME [64]. Unlike exoplanet events, whose lightcurves span ~10 days, BH events span several hundreds of days (note the x-scale below is years, while in the plots above it is days).

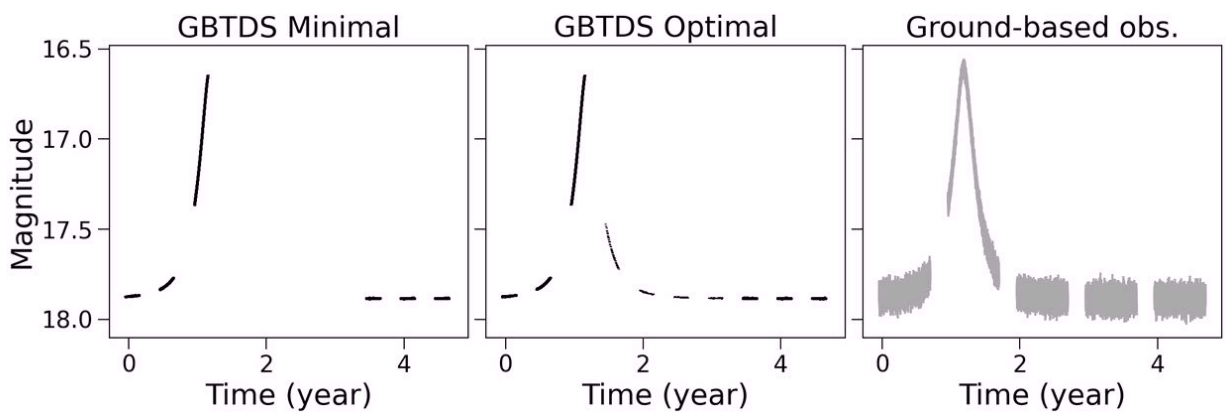

Roman's photometric precision is much better than ground-based surveys. However, **due to the large ~2.5 year gap between Bulge windows in the notional GBTDS design, most BH lightcurves will not be fully observed. Filling in the gaps will significantly help characterization.** Finally, ground based coverage will help by filling in Roman's bulge windows, which will be critical to issues like microlensing degeneracies and model mismatches.

# Figure 5: Astrometric microlensing, now vs. anticipated with Roman

Astrometric microlensing signal of a 9.4M$_\odot$ BH with an Einstein radius of 2 mas (corresponding to a maximum astrometric shift of ~0.7 mas) as would be detected by Roman. The simulated observations (black) assume 1 mas astrometric uncertainties for each individual exposure [65].

Left: astrometric signal as observed with the minimal/notional GBTDS. Right: astrometric signal as observed with the optimal GBTDS. During the 6 exoplanet-focused seasons, 4 days of observations at 15 min cadence are stacked to achieve ~0.05 mas precision; during the additional 4 seasons, 12 days of observations at 1 day cadence are stacked to achieve ~0.3 mas precision. **Filling in the gap is critical to detecting the maximal astrometric signal, which is needed to measure lens masses and velocities.** Including the 4 Bulge seasons increases the astrometric duty cycle by 60%.

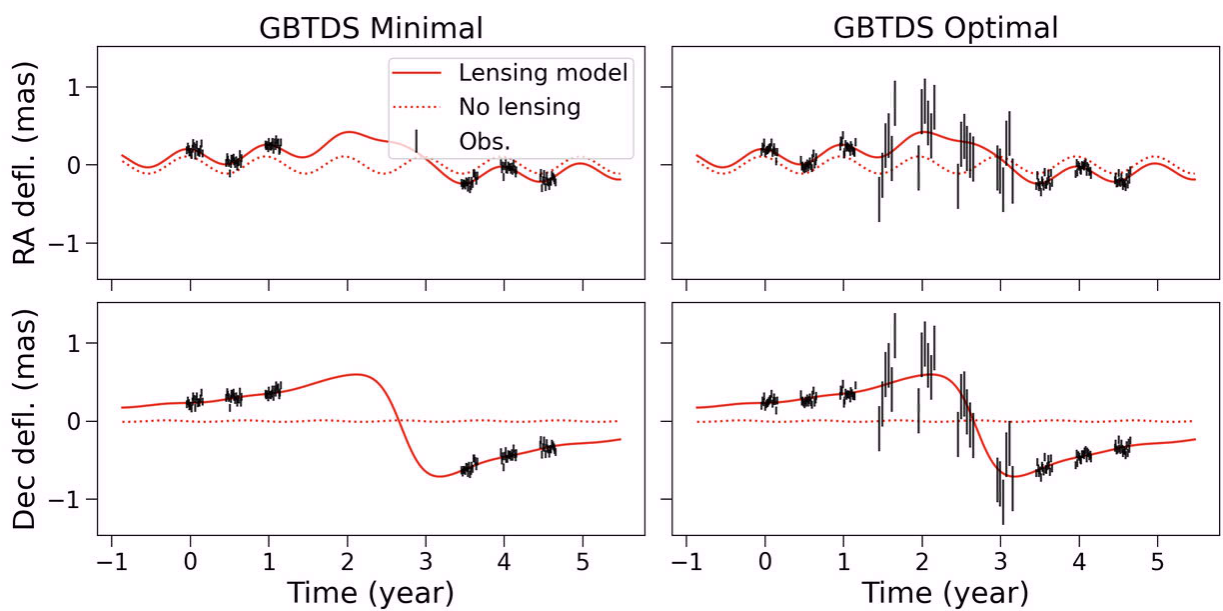

Note: the wiggles in the non-lensing model are due to parallax (which primarily shows up in the RA toward the Bulge).

We contrast this to current capabilities achievable using follow-up methods, where each target has <10 astrometric epochs and astrometric precisions ~0.3 mas. *Left*: Keck 10-m telescope equipped with laser guide star adaptive optics. *Right*: HST. Figures adapted from [25, 41].

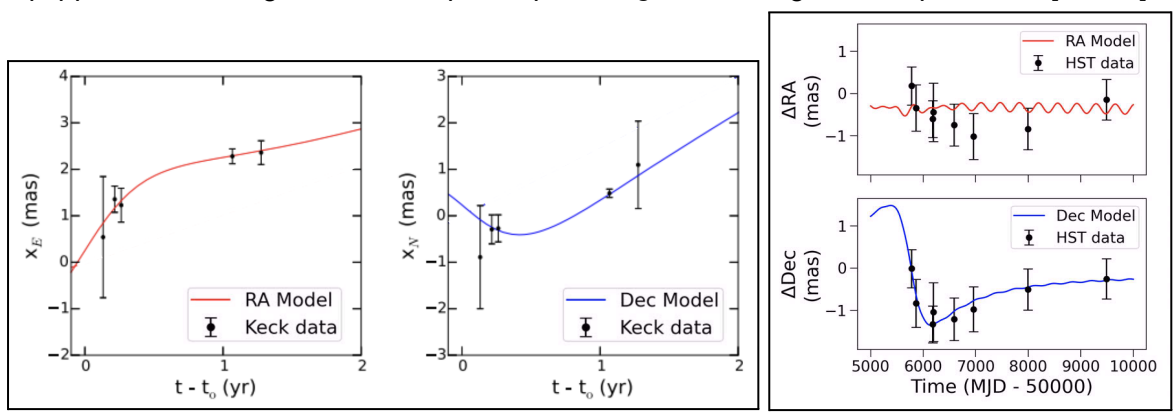

# Figure 6: Astrometric precision vs. number of detectable black holes

The number of BHs characterizable by Roman, as a function of the stacked astrometric precision. The y-axis assumes 300 total BHs in the Roman survey, as described on page 5. The bottom x-axis shows the maximum astrometric shift signal of the microlensing event. The top x-axis shows the corresponding precision needed to measure the astrometric signal; we assume the precision needed is 3 times smaller than the maximum signal[1].

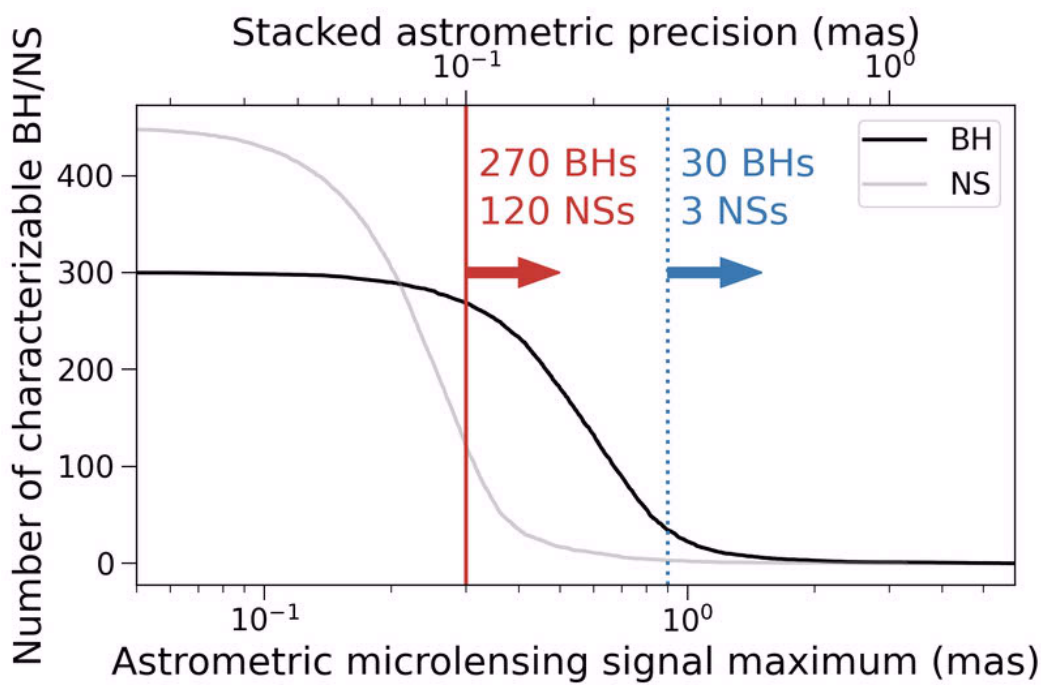

**Roughly 270 BH microlensing events have maximum astrometric shifts of at least 0.3 mas, and these can all be characterized if the astrometric precision is 0.1 mas** (red line and arrow). If the astrometric precision is instead 0.3 mas, then only about 30 of the BH microlensing events can be characterized (blue dotted line and arrow); this is still an order of magnitude more than the current sample size of 1 isolated BH.

For reference, we also show the corresponding curve for isolated neutron stars (NS). Many of the reasons that make isolated BHs astrophysically interesting are also true of NSs, in addition to the potential to answer questions related to the maximum NS mass and their equations of state. Although there are an order of magnitude more NSs than BHs in the Galaxy, the total lensing cross section of all NS is only about a factor of 1.5 larger than all BHs because of their lower mass. NSs also have astrometric shifts about 2 times smaller than BHs. Hence, **a significant sample of isolated NSs will not be detectable by Roman unless the optimal astrometric precision of 0.1 mas is achieved.**

---

[1] This depends on the sampling cadence and how the lightcurve is constrained; this estimate is based on the results of [25], which measured a 1.5 mas maximum shift with precisions of 0.3 mas, corresponding to a factor of 5, with only 10 total observations. A factor of 3 is conservative, given that there will be many more observations by Roman.

# Table 1: Observational strategy for the GBTDS

We assume that for the GBTDS to meet its core exoplanetary science goals, there must be at least 6 x 60 day seasons, with a cadence of at least 15 minutes, and observations during the first two and last two available Bulge window, over ~2 deg$^2$ (see Appendix A for a summary).

In the table below, green boxes denote strategies the same as the notional GBTDS; yellow boxes denote minor additions and changes to the notional GBTDS; orange boxes denote strategies not specified by the notional GBTDS.

| Strategy/parameter | Minimal | Optimal |
|---|---|---|
| Depth of each epoch | Same as notional | |
| Number of epochs and time between individual epochs | Same as notional: 6 seasons of 60 days, with continuous 15 minute sampling | 10 seasons (for the 4 non-exoplanet focused seasons, average ~1 day sampling cadence); 72 day seasons |
| Temporal baseline, first to last epoch | Same as notional: maximal, i.e. require observations in the first two and last two Bulge seasons | |
| Uniformity of time between individual epochs | Same as notional: Uniformity not crucial as there will be unavoidably large gaps by nature of the observatory; as continuous as possible coverage during available windows is the key. | |
| Final co-added depth | Same as notional | |
| Total survey area | Same as notional: ~2 deg$^2$ | As many fields as can be sustained where the cadence does not drop below ~ 1 day on average. |
| Locations of surveyed areas | Same as notional: regions of relatively low extinction in the Bulge | Inclusion of field(s) closer to b = 0 deg and Galactic Center where event rate is higher. |
| Specific filters | Same as notional: widest possible (F146) for photometric precision | |
| Number of filters | Same as notional: occasional (~ twice per season) observations in another filter to do CMDs, but not regularly as observations in the narrower filters will have worse photometric/astrometric precisions. | |
| Subpixel dithering | See WFIRST Astrometry Working Group white paper [65] | |
| Large gap dithers | See WFIRST Astrometry Working Group white paper [65] | |
| Ground-based coverage | Observations ~ 1/night, during all times the Bulge fields are visible from Earth | Observations ~ 10/night, during all times the Bulge fields are visible from Earth |
| Astrometric precision (10 day average) | 0.3 mas (e.g. what is achievable with HST now) | 0.1 mas |

## Table 2: Synergies with other white papers and pitches

The following white papers and pitches describe in more detail some of the topics we mention briefly or in passing.

| Topics and synergies | References (*Roman CCS white papers and pitches marked with* *) |
|---|---|
| Characterizing isolated stellar-mass black holes with microlensing | [66*, 67*, 68*, 69*] |
| Stellar-mass BHs in binaries and other compact objects | [70*, 71*, 72*] |
| Observations at the Galactic Center | [67*, 72*, 73*, 74*] |
| Improved Roman coverage during Bulge seasons | [68*, 75*] |
| Coordinated observations with other facilities existing and upcoming (PRIME, Rubin, JASMINE, ULTIMATE-Subaru[2]) | [67*, 72*, 76*, 77*, 78] |
| Importance of astrometry | [65, 66*, 72*] |

Below is a list of reference material for the listed facilities:

PRIME: 1.8-m telescope that has just begun (2023) a NIR microlensing survey of the inner Galactic bulge to help design the observing strategy for Roman's exoplanet microlensing survey
- Website: http://www-ir.ess.sci.osaka-u.ac.jp/prime/index.html
- See also [64]

Rubin: 8.4-m telescope that will conduct a 10-year survey of the Southern sky beginning ~2025
- Website: https://rubinobservatory.org/

ULTIMATE-Subaru: next-generation wide-field ground-layer adaptive optics development project on the 8.2-m Subaru Telescope; wide field instrument first light ~2027
- Website: https://ultimate.naoj.org/english/index.html

JASMINE: planned M-class mission by ISAS/JAXA, launch planned ~2028
- Website: http://www.scholarpedia.org/article/JASMINE

---

[2] Short slide deck for reference ("Science Goal (2): ULTIMATE for Transient Universe"): https://ultimate.naoj.org/superirnet/WS2023_talk_online/20230323_ULTIMATE_science_goals_koyama_public.pdf

# Appendix: GBTDS exoplanet requirements and notional design

For reference, we provide a summary of the mandatory constraints on the GBTDS needed to achieve the exoplanetary microlensing science requirements; the points below are copied from [slides](#) by Karoline Gilbert (Mission Scientist, Roman Mission Office).

Microlensing needs:
- Monitor hundreds of millions of bulge stars continuously on a time scale of <15 minutes
- Minimum 60 day seasons
- Precise Relative Photometry
- Resolve main sequence source stars for smallest planets.
- Resolve unrelated stars for lens flux measurements.
- Longest possible time baseline for proper motion measurements

Bounding conditions for GBTD Survey:
- Cadence of repeat visits and S/N per visit must be sufficient for sensitivity to the chosen range of planet masses (0.1 - 10000 *MEarth) in the Science Requirements Document.
- Area/cadence trade should provide monitoring for a minimum of 600 sq-degree-days, distributed over 6 seasons.
- The duty-cycle for observations devoted to this survey must be greater than 80% during each season.
  - This includes time required for momentum unloading and station-keeping (~9 hours/month or ~1.25%) and any other mission overheads.

Scheduling considerations -GBTD Survey
- Want continuous coverage of a particular field for entire visibility period
  - ≤72 days, Spring and Fall (DRM is 62 days, Penny et al is 72 days)
- Visits at 15-minute cadence for the core survey
  - This does not preclude adding additional fields with a equal (or longer cadence)
  - Or increasing the cadence for one of the fields
- Longest possible total time baseline
  - accurate proper motions (broad science benefit)
  - maximizing separation of stars in lensing events

For reference, we also summarize the notional design of the GBTDS, as described at this [page](#):
- Survey area: 7 WFI fields (~ 2 deg$^2$ total)
- Survey cadence: 15 minutes
- Survey season duration: 62 days
- Number of seasons: 6 (3 at the beginning of survey, and 3 at the end of survey)
- Filter: F146 primary, bluest filter (every 12 hours)

Note that the maximum 72 day survey season duration is determined by when the Galactic Bulge is visible to Roman.

# References

1. A. Heger et al. "How Massive Single Stars End Their Life". 2003, ApJ, 591, 288

2. E. O'Connor & C. D. Ott. "Black Hole Formation in Failing Core-Collapse Supernovae". ApJ, 730, 70

3. T. Sukhbold et al. "Core-collapse Supernovae from 9 to 120 Solar Masses Based on Neutrino-powered Explosions". 2016, ApJ, 821, 38

4. R. A. Patton & T. Sukhbold. "Towards a realistic explosion landscape for binary population synthesis". 2020, MNRAS, 499, 2803

5. I. Mandel. "Estimates of black hole natal kick velocities from observations of low-mass X-ray binaries". 2016, MNRAS, 456, 578

6. S. Repetto et al. "The Galactic distribution of X-ray binaries and its implications for compact object formation and natal kicks". 2017, MNRAS, 467, 298

7. P. Atri et al. "Potential kick velocity distribution of black hole X-ray binaries and implications for natal kicks". 2019, MNRAS, 489, 3116

8. G. Fragione et al. "Impact of Natal Kicks on Merger Rates and Spin-Orbit Misalignments of Black Hole-Neutron Star Mergers". 2021, ApJL, 918, L38

9. T. A. Callister et al. "State of the Field: Binary Black Hole Natal Kicks and Prospects for Isolated Field Formation after GWTC-2". 2021, ApJ, 920, 157

10. S. Wellstein & N. Langer. "Implications of massive close binaries for black hole formation and supernovae". 1999, A&A, 350, 148

11. P. Eggleton. "Evolutionary Processes in Binary and Multiple Stars". 2006, Cambridge, UK, Cambridge University Press, ISBN 0521855578

12. H. Sana et al. "Binary Interaction Dominates the Evolution of Massive Stars". 2012, Science, 337, 444

13. N. Ivanova et al. "Common envelope evolution: where we stand and how we can move forward". 2013, Astron. Astrophys. Rev., 21, 59

14. F. X. Timmes et al. "The Neutron Star and Black Hole Initial Mass Function". 1996, ApJ, 457, 834

15. M. Samland. "Modeling the Evolution of Disk Galaxies. II. Yields of Massive Stars". 1998, ApJ, 496, 155

16. E. Agol et al. "Finding Black Holes with Microlensing". 2002, ApJ, 576, L131

17. R. P. Fender et al. "The closest black holes". MNRAS, 430, 1538

18. G. Wiktorowicz et al. "Populations of Stellar-mass Black Holes from Binary Systems". ApJ, 885, 1

19. A. Olejak et al. "Synthetic catalog of black holes in the Milky Way". 2020, A&A, 638, A94

20. J. M. Corral-Santana et al. "BlackCAT: A catalogue of stellar-mass black holes in X-ray transients". 2016, A&A, 587, A61

21. T. A. Thompson et al. "A noninteracting low-mass black hole-giant star binary system". 2019, Science, 366, 637

22. K. El-Badry et al. "A Sun-like star orbiting a black hole". 2023, MNRAS, 518, 1057

23. S. Chakrabarti et al. "A non-interacting Galactic black hole candidate in a binary system with a main-sequence star". 2023, AJ, 166, 6

24. K. El-Badry et al. "A red giant orbiting a black hole". 2023, MNRAS, 521, 4323

25. C. Y. Lam et al. "An Isolated Mass-gap Black Hole or Neutron Star Detected with Astrometric Microlensing". 2022, ApJL, 933, L23

26. C. Y. Lam et al. "Supplement: "An Isolated Mass-gap Black Hole or Neutron Star Detected with Astrometric Microlensing" (2022, ApJL, 933, L23)". 2022, ApJS, 260, 55

27. K. C. Sahu et al. "An Isolated Stellar-mass Black Hole Detected through Astrometric Microlensing". 2022, ApJ, 933, 83

28. The LIGO, Virgo and KAGRA collaborations. "GWTC-3: Compact Binary Coalescences Observed by LIGO and Virgo During the Second Part of the Third Observing Run". 2021, arXiv:2111.03606v2 [gr-qc]

29. F. S. Broekgaarden et al. "Impact of massive binary star and cosmic evolution on gravitational wave observations I: black hole-neutron star mergers". 2021, MNRAS, 508, 5028

30. F. S. Broekgaarden et al. "Impact of massive binary star and cosmic evolution on gravitational wave observations - II. Double compact object rates and properties". 2022, MNRAS, 516, 5737

31. M. Chruślińska. "Chemical evolution of the Universe and its consequences for gravitational- wave astrophysics". 2022, Ann. Phys. (Berlin), 2200170

32. J. Binney et al. "Is Galactic Structure Compatible with Microlensing Data?". 2000, ApJ, 536, L99

33. M. Moniez. "Microlensing as a probe of the Galactic structure: 20 years of microlensing optical depth studies". 2010, Gen. Relativ. Gravit., 42, 2047

34. C. Wegg et al. "MOA-II Galactic microlensing constraints: the inner Milky Way has a low dark matter fraction and a near maximal disc". 2016, MNRAS, 463, 557

35. S. Chakrabarti et al. "Toward a Direct Measure of the Galactic Acceleration". 2020, ApJL, 902, L28

36. K. De. "Revealing the obscured and dynamic Galactic bulge with a wide area survey". Roman Science Pitch

37. W. Clarkson et al. "Stellar Proper Motions in the Galactic Bulge from Deep Hubble Space Telescope ACS WFC Photometry". 2008, ApJ, 684, 1110

38. R. Poleski et al. "The Optical Gravitational Lensing Experiment. The Catalog of Stellar Proper Motions toward the Magellanic Clouds". 2012, Acta Astron., 62, 1

39. J. Klüter et al. "Prediction of Astrometric-microlensing Events from Gaia eDR3 Proper Motions". 2022, AJ, 163, 176

40. A. del Pino et al. "GaiaHub: A Method for Combining Data from the Gaia and Hubble Space Telescopes to Derive Improved Proper Motions for Faint Stars". 2022, ApJ, 933, 76

41. J. R. Lu et al. "A Search For Stellar-mass Black Holes Via Astrometric Microlensing". 2016, ApJ, 830, 41

42. N. Kains et al. "Microlensing Constraints on the Mass of Single Stars from HST Astrometric Measurements". 2017, ApJ, 843, 145

43. J. Rybizki et al. "A classifier for spurious astrometric solutions in Gaia eDR3". 2022, MNRAS, 510, 2597

44. K. A. Rybicki et al. "On the accuracy of mass measurement for microlensing black holes as seen by Gaia and OGLE". 2018, MNRAS, 476, 2013

45. M. T. Penny et al. "Predictions of the WFIRST Microlensing Survey. I. Bound Planet Detection Rates". 2019, ApJS, 241, 3

46. A. Gould. "Measuring the Remnant Mass Function of the Galactic Bulge". 2000, ApJ, 535, 928

47. C. Y. Lam et al. "PopSyCLE: A New Population Synthesis Code for Compact Object Microlensing Events". 2020, ApJ, 889, 31

48. J. J. Andrews & V. Kalogera. "Constraining Black Hole Natal Kicks with Astrometric Microlensing". 2022, ApJ, 930, 159.

49. S. Rose et al. "The Impact of Initial-Final Mass Relations on Black Hole Microlensing". 2022, ApJ, 941, 116

50. B. Giesers et al. "A stellar census in globular clusters with MUSE: Binaries in NGC 3201". 2019, A&A 632, A3

51. B. Giesers et al. "A detached stellar-mass black hole candidate in the globular cluster NGC 3201". 2018, MNRAS, 475, L15

52. T. Shenar et al. "An X-ray-quiet black hole born with a negligible kick in a massive binary within the Large Magellanic Cloud". 2022, Nat. Astron., 6, 1085

53. W. M. Farr et al. "The Mass Distribution of Stellar-mass Black Holes". ApJ, 741, 103

54. F. Özel et al. "The Black Hole Mass Distribution in the Galaxy". ApJ, 725, 1918

55. P. G. Jonker et al. "The Observed Mass Distribution of Galactic Black Hole LMXBs Is Biased against Massive Black Holes". 2021, ApJ, 921, 131

56. S. Mao et al. "Optical Gravitational Lensing Experiment OGLE-1999-BUL-32: the longest ever microlensing event - evidence for a stellar mass black hole?". 2002, MNRAS, 329, 349

57. D. P. Bennett et al. "Gravitational Microlensing Events Due to Stellar-Mass Black Holes". 2002, ApJ, 579, 639

58. M. Dominik & K. C. Sahu. "Astrometric Microlensing of Stars". 2000, ApJ, 534, 213

59. M. A. Walker. "Microlensed Image Motions". 1995, ApJ, 453, 37

60. M. Miyamoto & Y. Yoshii. "Astrometry for Determining the MACHO Mass and Trajectory". AJ, 110, 1427

61. E. Hog et al. "MACHO photometry and astrometry". 1995, A&A, 294, 287

62. S. Mao. "Astrophysical applications of gravitational microlensing". 2012, Res. Astron. Astrophys., 12, 947

63. S. A. Johnson et al. "Predictions of the Nancy Grace Roman Space Telescope Galactic Exoplanet Survey. II. Free-floating Planet Detection Rates". 2020, AJ, 160, 123

64. I. Kondo et al. "Prediction of Planet Yields by the PRime-focus Infrared Microlensing Experiment Microlensing Survey". 2023, AJ, 165, 254

65. WFIRST Astrometry Working Group et al. "Astrometry with the Wide-Field Infrared Space Telescope". 2019, JATIS, 5, 044005

66. C. Lam et al. "Finding isolated black holes with the Galactic Bulge Time Domain Survey". Roman Science Pitch

67. D. Suzuki et al. "Black Hole Microlensing Survey with Roman and ULTIMATE-Subaru toward the Galactic Center". Roman Science Pitch

68. K. Sahu. "Detecting Isolated Stellar Mass Black Holes with Roman". Roman Science Pitch

69. S. Gaudi & D. Bennett. "The Roman Galactic Exoplanet Survey". Roman Science Pitch

70. A. Bahramian. "Demystifying hidden black holes and neutron stars in the Galactic Bulge". Roman Science Pitch

71. A. Bahramian et al. "X-ray binaries, cataclysmic variables and transients in the Galactic bulge". Roman CCS White Paper

72. D. Kawata et al. "Roman+JASMINE Galactic Center Astrometry Survey". Roman Science Pitch

73. S. Terry et al. "Transients at the Galactic Center with Roman". Roman Science Pitch

74. S. Terry et al. "The Galactic Center with Roman". Roman CCS White Paper

75. T. Sumi et al. "Search for quasars behind the Galactic bulge via variability". Roman Science Pitch

76. R. Street et al. "Optimizing the science return from simultaneous NIR and Optical Timeseries photometry in the Galactic Bulge". Roman Science Pitch

77. R. Street et al. "Maximizing science return by coordinating the survey strategies of Roman with Rubin, and other major facilities". Roman CCS White Paper

78. S. Gezari et al. "R2-D2: Roman and Rubin -- From Data to Discovery". AURA-commissioned white paper, arXiv:2202.12311v1 [astro-ph.IM]